# Quantitative Excited State Spectroscopy of a Single InGaAs Quantum Dot Molecule through Multi-million Atom Electronic Structure Calculations

Muhammad Usman[1, *], Yui-Hong Matthias Tan[1, *], Hoon Ryu[1], Shaikh S. Ahmed[2], Hubert Krenner[3], Timothy B. Boykin[4], and Gerhard Klimeck[1]

[*]Co-first authors, contributed equally

[1]School of Electrical and Computer Engineering and Network for Computational Nanotechnology, Purdue University, West Lafayette Indiana, 47906 USA
[2]Department of Electrical and Computer Engineering, Southern Illinois University at Carbondale, Carbondale, IL, 62901 USA
[3]Lehrstuhl für Experimentalphysik 1, Universität Augsburg, Universitätsstr. 1, 86159 Augsburg, Germany
[4]Department of Electrical and Computer Engineering, University of Alabama in Huntsville, Huntsville, AL, 35899, USA

Atomistic electronic structure calculations are performed to study the coherent inter-dot couplings of the electronic states in a single InGaAs quantum dot molecule. The experimentally observed excitonic spectrum by H. Krenner *et al*. [12] is quantitatively reproduced, and the correct energy states are identified based on a previously validated atomistic tight binding model. The extended devices are represented explicitly in space with 15 million atom structures. An excited state spectroscopy technique is applied where the externally applied electric field is swept to probe the ladder of the electronic energy levels (electron or hole) of one quantum dot through anti-crossings with the energy levels of the other quantum dot in a two quantum dot molecule. This technique can be used to estimate the spatial electron-hole spacing inside the quantum dot molecule as well as to reverse engineer quantum dot geometry parameters such as the quantum dot separation. Crystal deformation induced piezoelectric effects have been discussed in the literature as minor perturbations lifting degeneracies of the electron excited (P and D) states, thus affecting polarization alignment of wave function lobes for III-V Heterostructures such as single InAs/GaAs quantum dots. In contrast this work demonstrates the crucial importance of piezoelectricity to resolve the symmetries and energies of the excited states through matching the experimentally measured spectrum in an InGaAs quantum dot molecule under the influence of an electric field. Both linear and quadratic piezoelectric effects are studied for the first time for a quantum dot molecule and demonstrated to be indeed important. The net piezoelectric contribution is found to be critical in determining the correct energy spectrum, which is in contrast to recent studies reporting vanishing net piezoelectric contributions.

### *Introduction and Problem Background*

Quantum dots grown by strain-driven self-assembly attract much interest because they can be used to implement optical communication and quantum information processing [1, 2]. Recently, significant advancements in providing good stability, high experimental repeatability, electroluminescence, and controlled coupling have made III-V quantum dots a potential candidate for quantum computers. Based on single qubit (quantum bit) realization with an exciton in a single quantum dot [3], optical quantum gates also have been obtained with both an exciton and a biexciton within one dot [4]. Coupled quantum dot molecules (QDMs), therefore, are good candidates for spin-based [5], charge-based [6], and exciton-based [7, 8] qubits. It is desirable to excite single excitons with external electric fields. Vertically stacked QDMs have been suggested to host single or double qubits; these can then be controlled by optical pulses, electrical fields, or magnetic fields [7-11]. However, a very basic requirement necessary for realizing qubits in these structures is the prior achievement of entangled states between the two dots. In a recent experimental study [12], coherent quantum coupling in QDMs has been observed with different separation distances between two dots forming a QDM under the applied bias. However a detailed quantitative study for the identification of the states in the spectrum and their coupling under linear and quadratic piezoelectric effects has been missing. The theoretical study accompanied with the experiment [12] is based on a single band effective mass model and considered only two lowest conduction band ($E_1$ and $E_2$) energy levels and two highest valence band ($H_1$ and $H_2$) energy levels. Thus the figure 3(b) in the reference [12] plots only one anti-crossing ($E_1 \leftrightarrow H_1$) and compares it to the experimental measurement. Moreover, it did not take into account the effects of the nonlinear piezoelectricity because the nonlinear piezoelectric field polarization constants [24] were not available at the time of this study in 2005. Thus the published study did not include the symmetries of individual quantum dots nor did it model the energy state couplings quantitatively. In a quantum dot molecule, each quantum dot possesses a ladder of electronic energy levels which give rise to multiple anti-crossings due to the electrical field induced Stark shift. It is therefore essential that more than two electron and hole energy levels should be considered to identify the correct energy states in the experimental measurements.

In this work, we present an atomistic theoretical analysis of the experimental measurement including alloy randomness, interface roughness, atomistic strain and piezoelectric induced polarization anisotropy, and realistic sized boundary conditions, which we believe is essential to fully understand the complex physics of these multi-million atom nanostructures [16]. Both linear and nonlinear components of the piezoelectric field are included. The net piezoelectric field is found to be critical to resolve the symmetries and energies of the excited states. Our theoretical optical transition strengths match with the experimental quantum dot state coupling strengths. Furthermore, we sweep the externally applied electrical field from zero to 21kV/cm to probe the symmetry of the electron states in the lower quantum dot based on the inter-dot energy level anti-crossings between the lower and the upper quantum dots. Such 'level anti-crossing spectroscopic' (LACS) analysis [37] can be used for a direct and precise measurement of energy levels of one quantum dot placed near another quantum dot in the direction of the applied electrical field. It can also be helpful to quantitatively analyze 'tunnel coupling energies' of the electron and hole energy states through the inter-dot energy level resonances in the single quantum dot molecule configuration predicted for the 'quantum information technologies' [12]. Finally the spacing between the anti-crossings and electrical field induced stark shifts allow us to 'reverse engineer' the separation between the quantum dots inside the quantum dot molecule.

Quantum dot molecules grown by self-assembly are mechanically coupled to each other through long-range strain originating from lattice mismatch between the quantum dot and the surrounding buffer. Despite the symmetric shape of the quantum dots (dome or lens shape), the atomistic strain is in general inhomogeneous and anisotropic, involving not only hydrostatic and biaxial components but also non-vanishing shear components [16, 26, 27]. Due to the underlying crystal symmetry theoretical modeling of these quantum dot molecules requires realistically spatially large extended boundary conditions to capture the correct impact of long-range strain on the electronic spectrum typically extending 30 nm into the substrate and 20 nm on both sides in the lateral direction. A detailed analysis of strain induced coupling and shifts in band edges of identical and non-identical quantum dots has been presented in earlier publications [22, 36, 38].

*Past Studies of Piezoelectric Effects*

III-V Heterostructures such as InGaAs/GaAs quantum dots show piezoelectric effects originating from diagonal and shear strain components. The asymmetric piezoelectric potentials are critical in determining the correct anisotropy of electron P-states [23-29]. Past studies of quantum dot molecules [43] to investigate the effect of strain and inter-dot separations on entanglement of electronic states does not include piezoelectric effects. Recent studies based on atomistic pseudopotentials suggest for single InAs quantum dots [24, 25] that linear and quadratic piezoelectric effects tend to cancel each other, thus leading to an insignificant net piezoelectric effect. Another study based on a k.p continuum method [30] used experimental polarization constants (see first row in table 1) that overestimated the piezoelectric effect by 35% to 50% for coupled quantum dot systems [23]. This work, for the first time, based on realistically sized boundary conditions and a three-dimensional atomistic material representation, takes into account the correct atomistic asymmetry and built-in electrostatic fields. Linear and quadratic polarization constants (see table 1) recently calculated using ab initio calculations [23] are used to study the impact of piezoelectric effect on excitonic spectra. Our calculations on a QDM show a non-vanishing net piezoelectric effect which is critical in reproducing experimental excitonic spectra [12]. Such non-vanishing piezoelectric potentials in single quantum dots have also been predicted recently [26]. However, previous studies in the literature so far [23-30] describe piezoelectric effects as merely small perturbations that lift excited states (P and D -states) degeneracies (increase their splitting) and/or flip the orientation of wave function lobes. This work is the first evidence that inclusion of the piezoelectric effect is indispensible to reproduce an experimentally observed excitonic spectrum in a quantum dot molecule system and to identify the correct energy states. Furthermore, optical transition intensities are calculated to characterize dark and bright excitons and matched with experimentally obtained transition strengths.

### *NEMO 3-D Simulator*

In this letter, an experimentally observed optical spectrum [12] is reproduced and the excitonic states are identified using the NanoElectronic MOdeling tool (NEMO 3-D) [13-15]. NEMO 3-D enables the atomistic simulation and computation of strain and electronic structure in multi-million atoms nanostructures. It can handle strain and electronic structure calculations consisting of more than 64 and 52 million atoms, corresponding to nanostructures of $(110 \text{ nm})^3$ and $(101 \text{ nm})^3$, respectively [14, 15]. Strain is calculated using an atomistic Valence Force Field (VFF) method [18] with anharmonic corrections [31]. The electronic structure calculation is performed using a twenty band $sp^3d^5s^*$ nearest neighbor empirical tight binding model [17]. The tight binding parameters for InAs and GaAs have been published previously and are used without any adjustment [17]. The bulk-based atom-to-atom interactions are transferred into nano-scale devices where no significant bond charge redistribution or bond breaking is expected and strain is typically limited to around 8%. The strain and electronic structure properties of alloys are faithfully reproduced through an explicit disordered atomistic representation rather than an averaged potential representation. The explicit alloy representation also affords the ability to model device-to-device fluctuations, which are critical in today's devices. For realistic semi-conducting nano-scale systems our tight binding approach, employed in NEMO 3-D, has been validated quantitatively against experimental data in the past through the modeling of the Stark effect of single P impurities in Si [19], distinguishing P and As impurities in ultra-scaled FinFET devices [20], the valley splitting in miscut Si quantum wells on SiGe substrate [21], sequences of InAs quantum dots in InGaAs quantum wells [16], and optical properties of single and bilayer quantum dots [44].

### *Simulated Geometry*

Figure 1(a) shows the simulated geometry, which consists of two vertically stacked lens shaped $In_{0.5}Ga_{0.5}As$ quantum dots separated by a 10nm GaAs buffer. As indicated in the experiment [12], the modeled upper quantum dot is larger in size (*Base=21nm*, *Height=5nm*) as compared to the lower quantum dot (*Base=19nm*, *Height=4nm*). In the lateral dimensions, the GaAs buffer size is set to 60nm with periodic boundary conditions. The modeled GaAs substrate is 30nm deep and the lattice constant is fixed at the bottom. A GaAs buffer with large lateral depth has been used to correctly capture the impact of long range strain and piezoelectric effects which is critical in the study of such quantum dot devices [13, 14, 16, 26, 27]. The quantum dots are covered by another 30nm GaAs capping layer where atoms are allowed to move at the top layer subject to an open boundary condition. The electronic structure calculation is conducted over a smaller domain embedded within the strain domain using closed boundary conditions. Since the electronic states of interest are closely confined inside the quantum dots a smaller electronic domain size is sufficient to model the confined electronic states. The strain domain comprises a volume of ~15 million atoms, and the electronic domain a volume of ~9 million atoms. In accordance to the experiment, a static external electric field ($\vec{F}$) is applied in [001] growth direction and varied from zero to 23kV/cm.

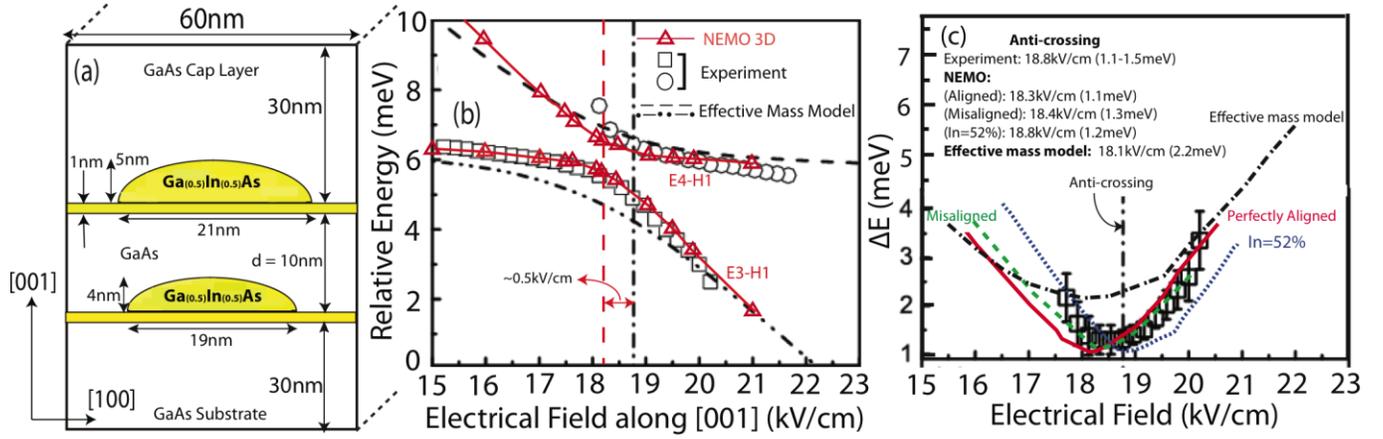

Figure 1: (a) Model system consisting of two lens shaped In$_{0.5}$Ga$_{0.5}$As quantum dots vertically stacked and separated by a 10nm GaAs buffer as described in the experiment. Both quantum dots are placed on 1nm thick In$_{0.5}$Ga$_{0.5}$As wetting layers. Substrate and cap layer thicknesses are 30nm. (b) NEMO 3D excitonic spectra (red triangles) for perfectly aligned quantum dots are compared with experimental measurement (black circles and squares) [12] and effective mass calculation [12] (dotted lines) [12]. The NEMO 3-D calculations match experiment quantatively and give a much better estimate of tunnel coupling energy than the effective mass model [12]. (c) Difference energy of excitons ($E_3$,$H_1$) and ($E_4$,$H_1$) in (b) is compared for various cases. Black squares with error bars are from experimental data. Solid line (red) is from NEMO 3-D structure in (a). Broken line (green) is for NEMO 3-D where the upper quantum dot in (a) is shifted to the right by 0.5nm. Dotted line (blue) is from NEMO 3-D with In$_{0.52}$Ga$_{0.48}$As quantum dots. Broken line with dots (black) is from the effective mass calculations [12]. A quantitative match of NEMO 3-D with experiment is evident. Small variations in quantum dot location and alloy composition insignificantly change the electrical field of the anticrossing and barely influence the exciton energy difference.

We mention here that based on the information provided by H. Krenner *et al*. [12, 42] regarding Schottky contacts and varying doping profiles of the experimental QDM structure, we estimated built-in electric fields of ~30kV/cm using self-consistent Poisson-Schrodinger calculations. In accordance to the experiment, the applied electric field as shown in our figures is referred to the flat-band voltage. Thus, for an applied field of $\vec{F} = 0$, the cumulative electric field including built-in fields in the quantum dot molecule is roughly zero (+/-0.5kV/cm).

## *Match with Experiment – Experimental Emission is from Excited States*

Figure 1(b) plots the excitonic energies as a function of applied bias. The curves indicated by circle and square data points are from experimentally obtained Photoluminescence measurements [12]. The measurements identify two bright excitonic emissions forming a tunable, coherently coupled quantum system. The triangle data points are from NEMO 3-D simulations. The excitonic spectra calculated here are based on a simple energy difference of the single electron and hole eigen energies. The charge to charge interaction will reduce the optical gap by around 5meV which we are ignoring in our calculations. We mention here that the experimental excitonic emission spectrum [12] was obtained through micro-photoluminescence experiments at low temperatures. A HeNe laser was used for excitation. In the experimental measurements, the excitation density (Pexc~ 2.5Wcm$^{-2}$) was kept low to ensure only generation of neutral single exciton species. We therefore conclude, that the experimentally observed excitonic emissions solely stem from neutral excitons and therefore calculations based on single electron and hole eigen energies are sufficient in understanding the experimental measurements.

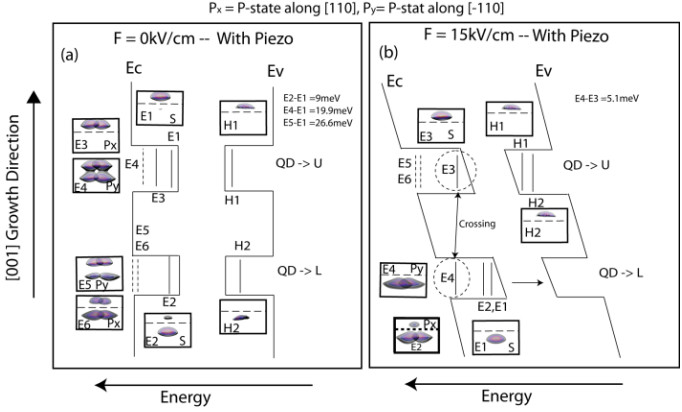

Figure 2: (a) Schematic lowest conduction ($E_c$) and highest valence band ($E_v$) edges with piezoelectric potential but zero external electrical field. Wave function plots are inserted for each energy state as small insets. The dotted lines separate the upper and lower quantum dots. Ground hole and electron states reside in the upper quantum dot because of its larger size [22, 36, 38]. (b) Schematic lowest conduction ($E_c$) and highest valence band ($E_v$) edges with piezoelectric potential and with 15kV/cm applied electrical field. Wave function plots are inserted for each energy state as small insets. The dotted lines separate the upper and lower quantum dots. Arrows are marked to show the tilting of band edges and the directions of movement of energy states when the electrical field is applied. The electrical field pushes $E_1$ and $E_2$ to the lower quantum dot, causing excitons $(E_1,H_1)$ and $(E_2,H_1)$ to be optically inactive. At 15kV/cm, $E_3$ makes a direct exciton with $H_1$ and hence will be optically active. Further increase in electrical field strength beyond 15kV/cm will push $E_3$ to the lower quantum dot and $E_4$ to the upper quantum dot resulting in an anti-crossing between $E_3$ with $E_4$ as observed in the experimental measurement [12].

Based on the simulation results, two excitons $(E_3,H_1)$ and $(E_4,H_1)$ are identified to match the experiment. Figure 1(c) compares the calculation of the exciton splitting $\Delta E = (E_4,H_1)-(E_3,H_1)$ obtained from NEMO 3-D with the experiment and a single band effective mass calculation [12]. The splitting at the anti-crossing point ($\Delta E_{min}$), referred to as the "tunneling coupling energy" [12] or the "anti-crossing energy" [37] is found to be ~1.1meV, which closely matches the experimental value of 1.1-1.5meV. On the other hand, the effective mass model significantly overestimates the tunneling coupling energy, predicting a value of ~2.2meV. Quantum dot molecules grown by self-assembly processes are neither perfectly aligned vertically [34], nor can the 'In' fraction of the quantum dot material be precisely determined [35]. These parameters are subject to slight variations during self-organization of the quantum dot nanostructures. Theoretical studies using NEMO 3-D on horizontally misaligned quantum dots (lateral misalignment = 0.5nm) and slight variation in the 'In' fraction ($In_{0.52}Ga_{0.48}As$ quantum dots instead of $In_{0.5}Ga_{0.5}As$ quantum dots) show that these difficult to control experimental imperfections can shift anti-crossing points to slightly higher electrical field values. For example, the anti-crossing point is found to be at 18.5kV/cm for the misaligned quantum dots and at 18.7kV/cm for increased indium fraction concentration as compared to 18.3kV/cm for perfectly aligned $In_{0.5}Ga_{0.5}As$ quantum dots. The exciton tunnel coupling energies appear to be almost insensitive to such experimental variations. The theoretical values show a close quantitative match with the experimental value of 18.8kV/cm, regardless of the small variations in quantum dot alignment and small alloy composition variations.

### *"Schematic" Band Edge Diagram – E3-E4 Anti-crossing in the Range of Experimental Field*

Figure 2(a) shows the spatial distribution of the single particle energy states in a "schematic" band edge diagram with piezoelectric field effects at zero electrical field. The wave function plots are inserted corresponding to each energy state to indicate their spatial occupation inside the quantum dot molecule and their single atom or molecular character. If the energy and position of $H_1$ is kept fixed and used as center of the electrical field lever arm shift, the application of an external [001] electrical field will tilt the band edges such that the lower quantum dot energy levels will move down in energy. Since the center of the electrical field lever arm shift is set at the position of the upper quantum dot, the lower quantum dot energy levels will exhibit a strong shift in the energy whereas the upper quantum dot energy levels will experience only a small Stark shift. As the lower quantum dot energy levels are shifted down by the electrical field, they anti-cross with the energy levels of the upper quantum dot and hence give rise to resonances. Figure 2(b) shows the new distribution of the energy levels at the electrical field strength of 15kV/cm. At 15kV/cm it is quite evident that, the lowest two electron states ($E_1$ and $E_2$) have already moved into the lower quantum dot. The excited electron state $E_3$ located in the upper quantum dots creates a direct exciton ($E_3,H_1$) with $H_1$. The excited state $E_4$, is in the lower quantum dot and forms an indirect exciton ($E_4,H_1$) at $\vec{F}$ = 15kV/cm. It can be anticipated that with further increase in the electrical field >15kV/cm, the conduction band edge will be tilted further. This will result in a decrease in the energy of $E_4$. The energy state $E_3$ will, therefore, anticross with the energy state $E_4$. This turns 'off' the optically active exciton ($E_3,H_1$) and turns 'on' the optically inactive exciton ($E_4,H_1$) as observed in the experiment [12]. Our calculations shown in the figure 1(b) demonstrate the anti-crossing between the excitons ($E_3,H_1$) and ($E_4,H_1$) in the range of applied bias (15kV/cm to 23kV/cm). A comprehensive spectroscopy of the energy levels for the electrical fields spanning zero to 21kV/cm range is presented in figure 3.

### *Electronic Spectrum Spectroscopy --- $H_1$ is taken as reference*

Figures 3(a, b) plots the electrical field dependence of the lowest four conduction band energy levels (electron energy states $E_1$, $E_2$, $E_3$, and $E_4$) and the highest four valence band energy levels (hole energy states $H_1$, $H_2$, $H_3$, and $H_4$) for the experimental quantum dot molecule geometry under study with (a) and without (b) piezoelectric fields. The [001] electrical field magnitude is varied from zero to 21kV/cm. The top most valence band (hole ground) state $H_1$ resides in the upper quantum dot at zero applied electrical field due to the larger size of the upper quantum dot and the dominance of the heavy hole (HH) band under these strain conditions [22, 36, 38]. The reference for the electrical field shift 'lever arm' is set to the top most valence band energy level $H_1$ to keep it fixed at its zero electrical field value. All the other energy levels are referenced to $H_1$.

### *Spectroscopy of Conduction Band States --- Three Anti-Crossings*

As shown in the figure 3(a), at the electrical field $\vec{F}$=0, the ground electron state $E_1$ is in the upper quantum dot due to its larger size [22, 36]. $E_2$ is in the lower quantum dot, whereas $E_3$ and $E_4$ are in the upper quantum dot and show $P_x$, $P_y$ like symmetry. The same position and character of the electronic states at $\vec{F}$=0 is also shown in the figure 2(a) using a schematic band edge diagram. The wave function plots are inserted corresponding to each energy state to indicate their spatial occupation inside the quantum dot molecule and their single atom or molecular character. As described earlier, the application of the external electrical field tilts the conduction bands, pushing the lower quantum dot states to the lower energies (see also figure 2(b)). On their way down to lower energies, the lower quantum dot electron energy states anti-cross with the upper quantum dot electron energy states and exhibit molecular like character. An anti-crossing results from the resonance of one quantum dot energy states (electron or hole) with the other quantum dot energy states when an external electrical field is applied. At the resonance, the electron states become spatially delocalized over the two quantum dots and form bonding and anti-bonding like molecular states. The hole states remain primarily localized in the separate dots. The separation between the anti-crossings provide a direct measurement of the energy levels of the lower quantum dot and hence enable 'reverse engineering' to determine the quantum dot dimension and symmetry. Such spectroscopic probing of the hole energy states in a quantum dot molecule has been done recently [37] and referred to as 'level anti-crossing spectroscopy'.

Figure 3(a) shows the distribution of three anti-crossings ($E_1\leftrightarrow E_2$, $E_2\leftrightarrow E_3$, and $E_3\leftrightarrow E_4$) between the electron energy levels. At $\vec{F}$=0, $E_1$ is the upper quantum dot and $E_2$ is the lower quantum dot. The first anti-crossing $E_1\leftrightarrow E_2$ results from the resonance of two s-type electronic states $E_1$ and $E_2$ when the electrical field is in the range of ~5kV/cm. The anti-crossing energy for $E_1\leftrightarrow E_2$ is ~5.9meV. For the electrical fields between 6kV/cm and 15kV/cm, $E_1$ resides in the lower quantum dot and $E_2$ resides in the upper quantum dot. The increase in the electrical field up to ~15kV/cm results in a second anti-crossing E2↔E3 between s-type upper quantum dot state ($E_2$) and p-type lower quantum dot state ($E_3$). The anti-crossing energy for this s-p anti-crossing (~0.7meV) is much smaller than the previous s-s anticrossing energy. This anti-crossing involves states that are higher in energy as measured from the relative bottom of the quantum well. The barrier height that separates/couples the states is noticeably smaller. As such the reduced coupling strength is at first unintuitive. The magnitude of the anti-crossing energy depends on the wave function overlap and the symmetries of the states involved [37]. The s-p anti-crossing energy will be smaller than the s-s anti-crossing energy because of the different spatial symmetries of the s- and p-states and smaller overlap between these two orbitals. Beyond $\vec{F}$=15kV/cm, both $E_1$ and $E_2$ are the atomic states confined in the lower quantum dot. Further electric field induced shift will bring $E_4$ (lower quantum dot state) into resonance with $E_3$ (upper quantum dot state) and exhibit a third anti-crossing $E_3 \leftrightarrow E_4$ between s-type upper quantum dot state and p-type lower quantum dot state. The anti-crossing energy for $E_3 \leftrightarrow E_4$ is ~1.1meV. Since the experiment [12] was performed for the electrical field varying from 15kV/cm to 21kV/cm, the experimentally observed anti-crossing (see figure 1(b)) is only E3↔E4.

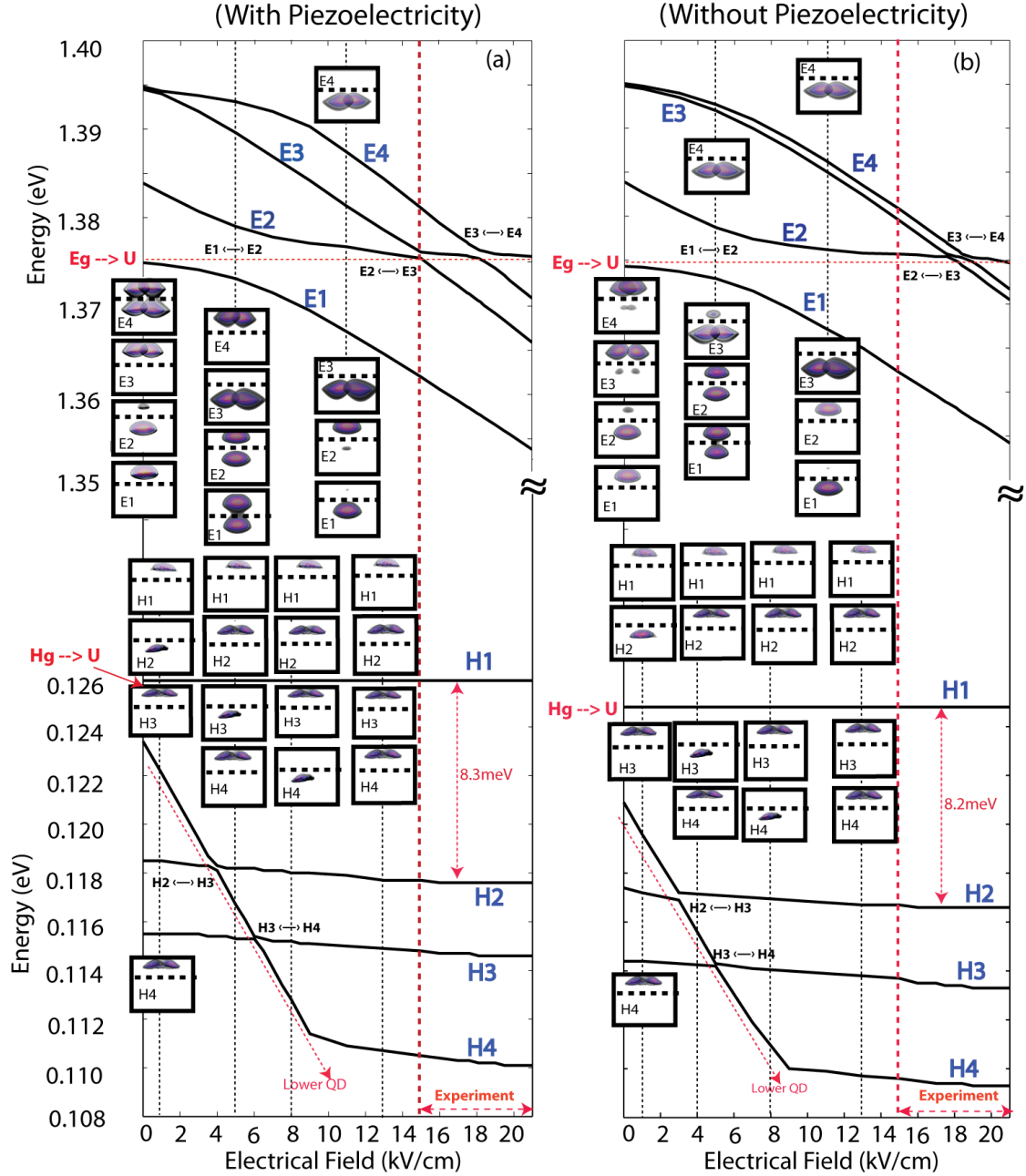

Figure 3: (a, b) Plot of first four electron and hole energy levels as a function of applied [001] electrical field. The electrical field is varied from zero to 21kV/cm. The $H_1$ energy is taken as the center of the electrical field lever arm so that its energy remains fixed at its zero electrical field value. Wave functions are inserted as small insets at critical values of electrical field to highlight the spatial position of the energy states inside the quantum dot molecule. The dotted lines in the wave function insets are marked to guide the eyes and separate the upper and lower quantum dots. Both the electron and the hole energy levels of the lower quantum dot exhibit anti-crossings with the energy levels of the upper quantum dot. The sequence of anti-crossings reveals the electron and hole level structure of the lower quantum dot and provide a direct measurement of the energy states in the lower quantum dot. The regions of the experimental electrical fields (from 15kV/cm to 21kV/cm) are zoomed-in in the figure 3, parts c and d.

*Spatial Separation of Electron and Hole States by Stark Shift Analysis*

The strength of the electrical field induced shift in the lower quantum dot energies depends on the separation between the quantum dots as the center of the electrical field lever arm shift is fixed at $H_1$ in the upper quantum dot. The slope of this shift calculated from figure 3(a) is $\Delta E/\Delta \vec{F}$ ~ -1.33 meV kV$^{-1}$ cm. This predicts the spatial separation between $H_1$ and the lower quantum dot electron states to be around 13.3nm. Since the wetting layer separation is 10nm, the center-to-center distance between the quantum dots is 11.5nm. The small difference of 13.3-11.5=1.8nm between the predicted value and the center-to-center distance of the quantum dots is due to spatial separation between the electron and hole states in the quantum dot. The hole states tend to reside closer to the top of the quantum dot and the electron states tend to reside closer to the bottom of the quantum dot [33]. This is also consistent with the slope of upper quantum dot electron energy shift which is around $\Delta E/\Delta \vec{F}$ ~ -0.167 meV kV$^{-1}$ cm. This corresponds to an intra-dot electron-hole spatial separation of 1.67nm, close to the value of 1.8nm mentioned above.

*Quantum Dot Spacing by Stark Shift Analysis*

The spectrum shown in the figure 3(a) clearly shows the spectroscopic probing of the lower quantum dot electron states by the upper quantum dot state $E_1$. This technique is very useful to determine the geometry parameters of the molecular quantum dot. For example, if we do not know the separation between the quantum dots, the $E_1 \leftrightarrow E_2$ anti-crossing values of $\vec{F}$ and slope of the energy shifts can be used to estimate the separation between the quantum dots. In the quantum dot molecule under study, $E_4$-$E_1$ is ~19.9meV as shown in the figure 2(a) at $\vec{F}$=0. The anti-crossing between these two states occurs at ~18.3kV/cm. This gives approximately 10.87nm separation between the two quantum dots which is very close to the value of ~10nm provided by the TEM measurements [12]. Thus the spacing between the anti-crossings and the electrical field induced stark shift allow us to 'reverse engineer' the separation between the quantum dots inside the quantum dot molecule.

*Optically Active States can be Referred as the Upper Quantum Dot Ground States*

The ground hole state $H_1$ resides in the upper quantum dot as shown in figure 2 in this field direction regardless of the field strength. We therefore mark this state as the ground hole state of the upper quantum dot: $H_g \rightarrow U$. The electron ground state of the upper quantum dot is marked as $E_g \rightarrow U$ and shown as horizontal dotted line in the figure 3(a). The optically active exciton will always be comprised of these two ground states of the upper quantum dot. It should be pointed out that although $E_g \rightarrow U$ is referred to as the upper quantum dot ground state; however this energy state undergoes several anti-crossings with the lower quantum dot states and exhibit atomic and molecular characters at the increasing values of the electrical fields. More importantly at the anti-crossing point E3↔E4, the $E_g$→U state hybridizes with the lower quantum dot P-state. This is very critical because it determines the importance of the piezoelectric fields which has stronger effect for electron P-states as compared to electron s-states as will be described in details later.

*Valence Band Spectroscopy --- Hole Anti-crossings*

Figure 3(a) also demonstrates the distribution of the hole energy level anti-crossings as a function of the applied electrical field. The electrical field pushes the hole energy levels of the lower quantum dot further to the lower energy values and thus provide a spectroscopic image of the upper quantum dot hole energy levels as they anti-cross between the two quantum dots. Since $H_1$ is in the upper quantum dot at $\vec{F}$=0, so it remains in the upper quantum dot for the applied electrical field direction and the valence band edge tilt. $H_2$ is in the lower quantum dot and as it moves down, it couples with upper quantum dot states and gives rise to the anti-crossings $H_2$↔$H_3$ and $H_3$↔$H_4$. The hole anti-crossing energies are ~0.2meV and ~0.1meV for $H_2$↔$H_3$ and $H_3$↔$H_4$ respectively. These hole anti-crossing energies are significantly smaller than the corresponding electron anti-crossing energies, indicating a much stronger hole localization. Notably in the range of the experimental electrical field from 15kV/cm to 21kV/cm, all the four hole levels are in the upper quantum dot and no anti-crossing is observed. This implies that the electron states in the lower quantum dot exhibit weak optical transition strengths with the first four hole energy levels and hence are not measured experimentally.

*Energy Spectrum Without and With Piezoelectricity in the Range of Experiment*

The comparison of the figures 3(a) and 3(b) highlights the impact of the piezoelectric fields on the energy spectrum of the quantum dot molecule under study. The piezoelectricity insignificantly impacts the hole energy spectrum. The hole energy levels exhibit only a small energy shift. Only minor changes are observed in the hole anti-crossings. The electron ground state energies of the two quantum dots namely $E_1$ and $E_2$ and their anti-crossing $E_1$↔$E_2$ is also shifted by a small amount. However the electron excited states ($E_3$, $E_4$) are significantly affected. Piezoelectricity increases the splitting between E3 and E4 and thus changes the position of the anti-crossing $E_2$↔$E_3$ on the electrical field axis. With piezoelectricity, $E_2$↔$E_3$ occurs outside the range of experimentally applied electrical field; without piezoelectricity on the other hand the anti-crossing $E_2$↔$E_3$ is pushed inside the range of the experimentally applied electrical field. This implies that without the piezoelectric field, there will be two anti-crossings in the range of the electrical field from 15kV/cm to 21kV/cm, which contradicts the experimental measurements as shown in the figure 1(b). The spectra for the electrical fields between 15kV/cm and 21kV/cm are enlarged in figure 3(c) and 3(d) to further explain the anti-crossings in this range of applied electrical field.

Figures 3 (c, d) depict the three exciton energies ($E_2,H_1$), ($E_3,H_1$), and ($E_4,H_1$) of the system with (c) and without (d) piezoelectric potential as a function of applied electrical field. $H_1$ hole as well as $E_2$, $E_3$, $E_4$ electron wave functions are depicted as insets for a few critical fields. If both the electron and hole states are in the same quantum dot, the resulting exciton is called a "direct exciton". It corresponds to an optically active state due to large electron hole overlaps. On the other hand, if the electron and hole states are in different quantum dots, then the exciton is called an "indirect exciton". It corresponds to an optically inactive state as the electron and hole overlap is negligible. Energy perturbations due to the quantum confined stark effect (QCSE) in the direct and indirect excitons under applied bias exhibit different slopes: $\Delta E_{exciton} = p.\vec{F}$, where $p$ is the dipole moment of the exciton, and $\vec{F}$ is the applied electrical field. The exciton dipole moment is defined as: $p$ = q.d, where 'q' is the electronic charge and 'd' is the electron-hole spatial separation. For a direct exciton, the electron and hole both are in the same quantum dot, and are separated by a very small distance (typically d ≤ 1nm). The resulting magnitude of the Stark shift is therefore very small. For an indirect exciton, the electron and hole are in different quantum dots and separated by a large spatial distance (d ≥ 10nm in our case). In this case, the dipole moment will be large and result in a strong magnitude of the Stark shift [33]. Figure 3 (c) and (d) depict the direct and indirect nature of the excitons over the range of applied bias. As the figures indicate, a strong (larger slope) and a weak (smaller slope) Stark effect correspond to an indirect and direct exciton, respectively.

### *With Piezoelectricity → One Anticrossing in the experimental data range*

The experimentally measured optical spectra (circles and triangles in figures 3(e) and 3(f)) show the relative strength of the two excitonic emissions 'A' and 'B'. At $\vec{F}$=15kV/cm, the excitonic emissions 'A' is bright and 'B' is dark. As the electrical field is increased, the intensity of 'A' becomes weaker and the intensity of 'B' gets stronger. The optical strengths of both excitons 'A' and 'B' become comparable at the electrical field value of 18.8kV/cm where these two excitons anti-cross. For the higher values of the applied electrical field, the intensity of 'A' quenches rapidly and 'B' becomes optically bright. We compute the transition rate intensities using Fermi's golden rule as the squared absolute value of the momentum matrix: $|<E_{2or3or4}|[\vec{n},\mathbf{H}]|H_1>|^2$, where **H** is single particle tight binding Hamiltonian, $E_{2or3or4}$ are electron states, $H_1$ is the top most valence band hole state, and $\vec{n}$ is position vector along the polarization direction of the incident light [32, 39-41]. Figure 3(e) compares the relative transition rate intensities $I_{(E_4,H_1)or(E_3,H_1)}/(I_{(E_4,H_1)} + I_{(E_3,H_1)})$, where $I_{(E_4,H_1)}$ and $I_{(E_3,H_1)}$ are the transition rate intensities of $(E_4,H_1)$ and $(E_3,H_1)$ respectively corresponding to the figure 3(a) in the experimental field range. Figure 3(f) compares the relative transition rate intensities $I_{(E_2,H_1)}orI_{(E_3,H_1)}orI_{(E_4,H_1)}/(I_{(E_4,H_1)} + I_{(E_3,H_1)} + I_{(E_2,H_1)})$, where $I_{(E_2,H_1)}$, $I_{(E_3,H_1)}$ and $I_{(E_4,H_1)}$ are the transition rate intensities of $(E_2,H_1)$, $(E_3,H_1)$ and $(E_4,H_1)$, respectively corresponding to the figure 3(b) in the experimental field range. Figure 3(e) compares the calculated optical strengths from NEMO 3D simulator including the piezoelectric effects with the experimental measurements. In this case, as the electric field is increased, the intensity of $(E_3,H_1)$ reduces rapidly, whereas the intensity of $(E_4,H_1)$ increases. Figure 2(c) plots the excitonic energies $(E_2,H_1)$, $(E_3,H_1)$ and $(E_4,H_1)$ as a function of the applied electrical field. Since $E_2$ remains in the lower quantum dot forming an indirect exciton with respect to $H_1$, $(E_2,H_1)$, in the range of the applied electrical field from 15kV/cm to 21kV/cm, its optical strength is very weak and is not measured in the experiment. The exciton $(E_3,H_1)$ is a direct exciton at 15kV/cm and results a strong optical peak. The increase in the electrical field results in the tunneling of $E_3$ from the upper quantum dot to the lower quantum dot, which changes the character of $(E_3,H_1)$ from an optically active state to an optically inactive state at $\vec{F}$~18.8kV/cm. In the vicinity of the anti-crossing point ($\vec{F}$ from ~17.9kV/cm to ~20.5kV/cm), two excitons $(E_3,H_1)$ and $(E_4,H_1)$ are tuned into resonance and exhibit close to equal magnitude of intensity. In that field range, the electron states ($E_3$ and $E_4$) are found to be delocalized over two quantum dots showing molecular like nature rather than representing well confined atomic states. Note that the relative intensities of the excitons in figure 3(e) calculated by our model closely follow the slopes of the experimental curves. NEMO 3-D is hence able to correctly capture the dynamics of controlled coupling under the resonance of excitons. The controlled coupling of quantum dots under external bias (gate voltage) is critical in the implementation of exciton qubits and may foster efforts in quantum information processing based on quantum dots.

### *Without Piezoelectricity → two Anticrossing*

Figure 3(f) compares the experimentally measured transition strengths ('A' and 'B') with the calculated transition strengths (solid lines) from the NEMO 3-D simulator without including the piezoelectric effects. The calculated transition strengths indicate three bright peaks and two anti-crossings. At $\vec{F}$=15kV/cm, the exciton $(E_2,H_1)$ is bright and the excitons $(E_3,H_1)$ and $(E_4,H_1)$ are dark. As the electrical field increases, first $(E_2,H_1)$ anti-cross with the exciton $(E_3,H_1)$. Further increase in the electrical field results in another anti-crossing between the

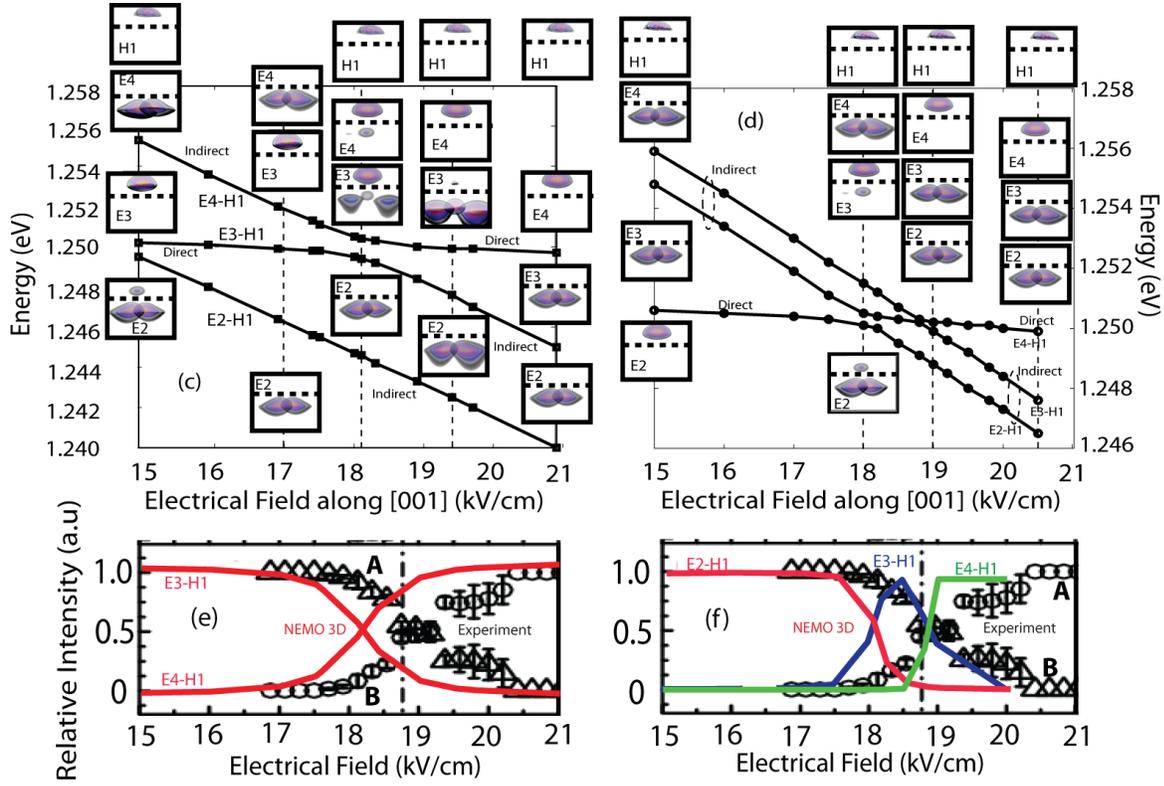

Figure 3 (c, d) Excitons $(E_2,H_1)$, $(E_3,H_1)$, and $(E_4,H_1)$ are shown as a function of [001] electrical field with (c) and without (d) piezoelectric effects. Insets show the wave function plots at (c) 15kV/cm, 18.2kV/cm, and 19.5kV/cm, (d) 15kV/cm, 18kV/cm and 19kV/cm. The dotted lines in the wave function insets are marked to guide the eyes and separate the upper and lower quantum dots. In the case of no piezoelectricity, two anti-crossings clearly mismatch the experimental measurement [12]. (e, f) Plots of the relative intensity of the interband optical absorption strengths as a function of the electrical field with (e) and without (f) the piezoelectricity. The solid lines are from NEMO 3-D calculations. Circles and triangles are from the experimental measurements [12]. It is clearly evident from the comparison of (e) and (f) that the piezoelectricity is critical to reproduce the experimentally measured optical spectra [12].

excitons $(E_3,H_1)$ and $(E_4,H_1)$. Figure 3(d) plots the corresponding excitonic energies as a function of the applied electrical field ($\vec{F}$). The direct and indirect nature of the excitons follows the optical transition strength patterns of figure 3(d). At $\vec{F}$=15kV/cm, the exciton $(E_2,H_1)$ is direct and the excitons $(E_3,H_1)$ and $(E_4,H_1)$ are indirect excitons. The first anti-crossing makes the exciton $(E_2,H_1)$ indirect and the exciton $(E_3,H_1)$ direct. The second anti-crossing makes the exciton $(E_3,H_1)$ indirect and the exciton $(E_4,H_1)$ direct. In short, the excitonic spectra without piezoelectricity as shown in the figures 3(d) and 3(f) result in three bright excitonic emissions, which disagree with the experimentally measured spectra.

Notably from the figures 3(c) and 3(d), the anti-crossings of the electron states in the range of the experimental measurement involve electron P-states of the lower quantum dot. Since the piezoelectric effects are more pronounced in the excited states [23-26, 28, 29, 38], the correct identification of optically active states in the experimental measurements is of crucial importance. If the experimental emissions came from the ground state, the piezoelectric effects would be of minor importance and merely a small correction in the excitonic energies. Figures 3(c) and 3(d) clearly show that including piezoelectric effects is important to obtain the correct resonance at the point of anti-crossing. Furthermore the field dependence of the optical intensity is also different between the two cases of with piezo and without piezo. We obtain the smoothly varying intensity variation as obtained by the experiment only when the piezoelectric fields are included [12].

## Piezoelectric Model

Piezoelectricity is calculated from electrical polarization originating from stressed crystals lacking inversion symmetry [29]. InGaAs/GaAs quantum dots are nanostructures with a lattice mismatch (~7%) which leads to long-range strain fields penetrating deep into the surrounding buffer [22, 27, 38]. The diagonal and off-diagonal (shear) strain components can result into a built-in piezoelectric polarization, which cannot be ignored. Bester et al. [24] and Ahmed et al. [27] outlined the importance of linear piezoelectric effect for single InAs quantum dots. Later, Bester et al. [25] highlighted the significance of the second order (quadratic) component of piezoelectricity. They concluded that the second order piezoelectric effect cancels the first order (linear) component, thus leading to a net piezoelectric effect significantly suppressed. Following their argument that the net piezoelectric effect and its impact on the excitonic spectrum is negligible, they claimed that it is better not to consider piezoelectric effects at all rather than including only the linear component of the piezoelectric effect. On the other hand, Schliwa et al. [28, 29] have indicated the importance of net piezoelectricity for some shapes and sizes of single quantum dots. The work presented here on quantum dot molecules clearly emphasizes that net piezoelectric effects cannot be neglected and are of critical importance in determining correct excitonic emission spectra in coupled dot system that involve occupation of excited states.

In our model, the piezoelectric potential is included into the single Particle Hamiltonian according to the recipe of reference [24, 29]. First both linear $\mathbf{P_1}$ (first order) and quadratic $\mathbf{P_2}$ (second order) polarizations are separately calculated from the strain tensor components (equations (1) and (2)) using already published polarization constants ($e_{14}$, $\beta_{114}$, $\beta_{124}$, $\beta_{156}$) (See table 1) [24, 25]. They are then added to calculate the total polarization $\mathbf{P}=\mathbf{P_1}+\mathbf{P_2}$:

$$\mathbf{P_1} = 2e_{14} \begin{pmatrix} \epsilon_{yz} \\ \epsilon_{xz} \\ \epsilon_{xy} \end{pmatrix} \quad (1)$$

$$\mathbf{P_2} = B_{114} \begin{pmatrix} \epsilon_{xx}\epsilon_{yz} \\ \epsilon_{yy}\epsilon_{xz} \\ \epsilon_{zz}\epsilon_{xy} \end{pmatrix} + B_{124} \begin{pmatrix} \epsilon_{yz}(\epsilon_{yy}+\epsilon_{zz}) \\ \epsilon_{xz}(\epsilon_{zz}+\epsilon_{xx}) \\ \epsilon_{xy}(\epsilon_{xx}+\epsilon_{yy}) \end{pmatrix} + B_{156} \begin{pmatrix} \epsilon_{xz}\epsilon_{xy} \\ \epsilon_{yz}\epsilon_{xy} \\ \epsilon_{yz}\epsilon_{xz} \end{pmatrix} \quad (2)$$

The divergence of the total polarization $\mathbf{P}$ is calculated over a rectangular mesh using a finite difference approach to calculate total charge density $\rho(\mathbf{r})$:

$$\rho(\mathbf{r}) = -\nabla.\mathbf{P} \quad (3)$$

Finally, the Poisson equation (4) is solved to calculate the piezoelectric potential $V_p(\mathbf{r})$, taking into account the position dependence of the static dielectric constant, $\varepsilon_s(\mathbf{r})$. The value of the dielectric constant for vacuum $\varepsilon_0$ is 8.85x10-12F/m. For InGaAs and GaAs materials, we used the relative dielectric constant values of $14.0\varepsilon_0$ and $12.84\varepsilon_0$, respectively.

$$\rho(\mathbf{r}) = \varepsilon_0 \nabla.[\varepsilon_s(\mathbf{r}) \nabla V_p(\mathbf{r})] \quad (4)$$

## Quadrupole Nature of the Piezoelectric Potentials

Figure 4(a) shows the first order (solid line with square), the second order (dashed line), and the sum of the first and second order (solid line with circle) piezoelectric potentials along the [001] direction through the center of the quantum dot molecule. It is evident that the net piezoelectric potential is nonzero inside the quantum dot and penetrates into the surrounding GaAs buffer. The reason for a nonzero net potential inside the quantum dot is that the quadratic (second order –> dashed line) component of the potential is significantly reduced for alloyed quantum dots (In$_{0.5}$Ga$_{0.5}$As) due to the following reasons: (1) Increasing the 'Ga' concentration reduces the diagonal strain components that determine the quadratic component, (2) The polarization constant $B_{124}$, which comprises a major portion of quadratic component is reduced in magnitude. The linear (first order –> solid line with square) component on the other hand increases mainly due to the large increase of $e_{14}$. The reduced quadratic component cannot cancel the effect of the already enhanced linear component, and hence the interior of the quantum dot will have a nonzero potential largely dominated by the linear component (see the solid line with circle). Outside of the quantum dot, the quadratic component is negligible, and only the linear component is dominant. As a result, the net piezoelectric effect inside and at the interfaces of the quantum dots is no longer negligible.

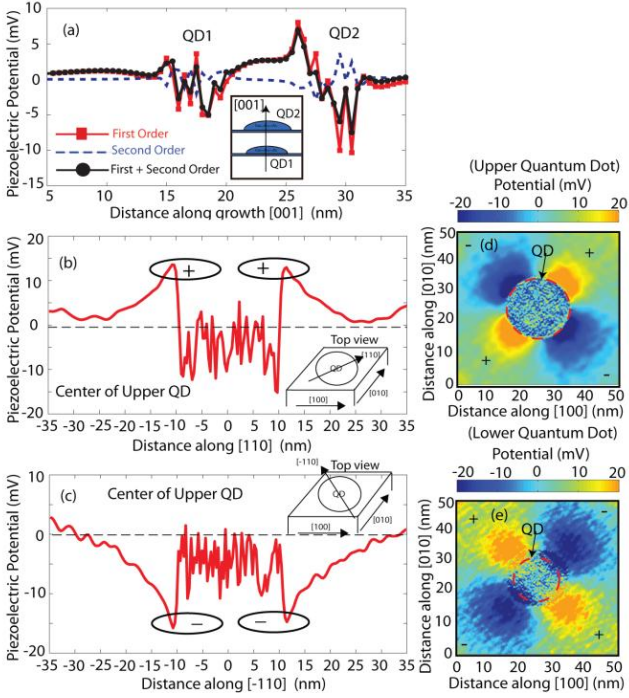

Table 1: Polarization constants for calculation of piezoelectric potential from reference [24]. The values for $In_{0.5}Ga_{0.5}As$ are obtained by linear interpolation between GaAs and InAs.

| Polarization Constant | Type of Component | GaAs (C/m$^2$) | InAs (C/m$^2$) | $In_{0.5}Ga_{0.5}As$ (C/m$^2$) |
|---|---|---|---|---|
| $e_{14}$ | Linear (Experiment) | -0.16 | -0.045 | -0.1025 |
| $e_{14}$ | Linear (Calculated) | -0.23 | -0.115 | -0.1725 |
| $B_{114}$ | Quadratic (Calculated) | -0.439 | -0.531 | -0.485 |
| $B_{124}$ | Quadratic (Calculated) | -3.765 | -4.076 | -3.9205 |
| $B_{156}$ | Quadratic (Calculated) | -0.492 | -0.12 | -0.306 |

Figure 4: (a) The first order (line with the box), second order (dashed line) and the sum of first and second order (line with the circle) piezoelectric potential are shown through the center of the quantum dot along the [001] growth direction. The net potential is clearly nonzero inside the quantum dots and at the interfaces of quantum dot and surrounding buffer. The second order piezoelectric effect is weak and cannot cancel first order effect inside the quantum dot. (b, c) Piezoelectric potential plotted through center of the upper quantum dot in the [110] and [$\bar{1}$10] directions. The origin of the x-axis is taken at the center of the quantum dot. The quadrupole nature of the potentials is clearly evident: the potential is positive at the interface of quantum dot and buffer along [110] direction, whereas it is negative at the interface of quantum dot and buffer along [$\bar{1}$10] direction. The random nature of potential inside the quantum dot region is a result of the random alloy configuration of $In_{0.5}Ga_{0.5}As$ quantum dot material. The quadrupole nature of the piezoelectric potential along diagonal directions strongly affects the splitting and orientation of electron P-states which are aligned along these directions because of atomistic asymmetry. (d, e) The two dimensional (xy) contour plots of the net piezoelectric potential at the base of the lower and upper quantum dots are shown as a function of distance along [100]- and [010]-axis. The quadrupole nature of the potential is clearly evident from the plots. Also, the potential is long range and penetrates deep inside the GaAs buffer. This imposes the necessity of large GaAs buffer in the lateral dimensions to fully incorporate the effect of potential in the electronic structure calculations.

Figures 4(b, c) show the net piezoelectric potential along the diagonal [110] and [$\bar{1}$10] directions through the center of the upper quantum dot. Figure 4(d, e) show the two dimensional contour plots of the net piezoelectric potentials about 1nm above the base of the lower and upper quantum dots. The quadrupole nature of the net piezoelectric potentials along the diagonal directions is clearly evident. The Piezoelectric potentials along [110] and [$\bar{1}$10] directions are large at the interfaces where strain is large, and penetrate deep inside the GaAs buffer [26, 27]. A large GaAs buffer is therefore not only required along [001] direction, but also in the plane of the quantum dot to fully capture the effect of the piezoelectric potential reaching 25-30nm deep into the GaAs buffer. The wide extension of the piezoelectric fields in the lateral plane also implies that the quantum dots that are closer

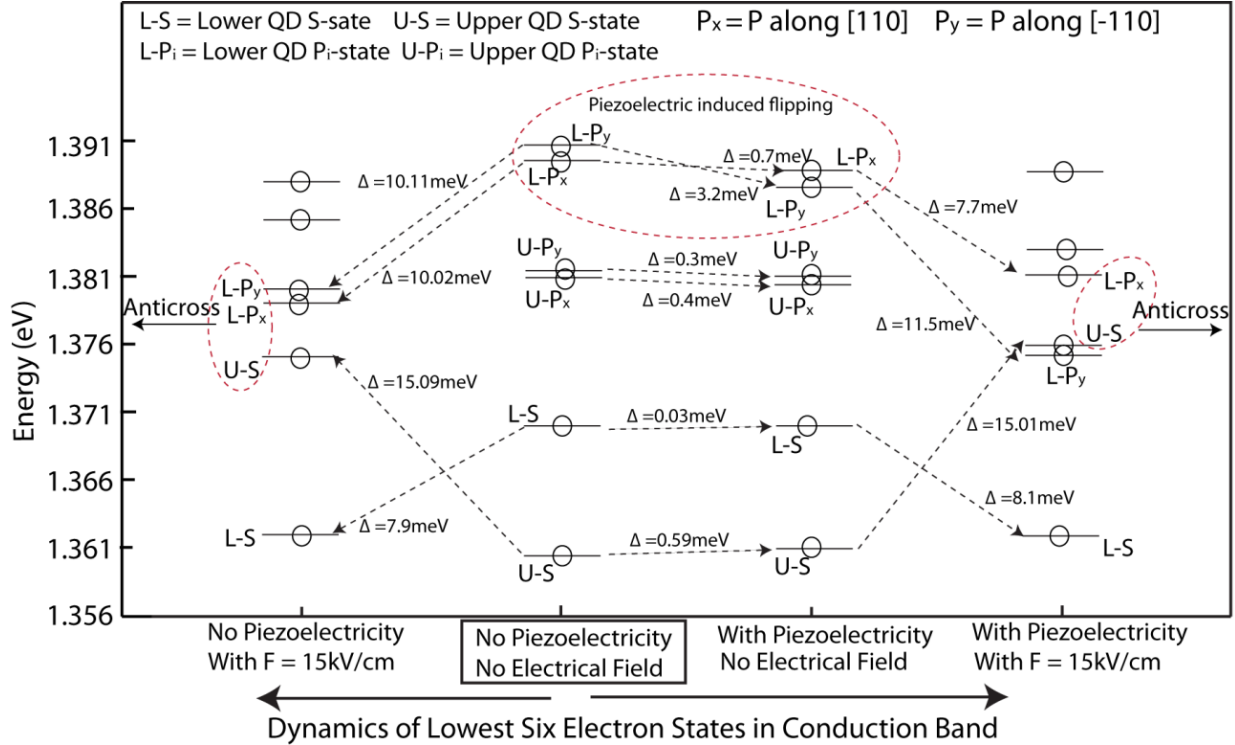

Figure 5: Flow diagram showing the ordering and position of first six electron states in the conduction band of quantum dot molecule. Starting from "No Piezoelectricity" case, right direction shows the inclusion of piezoelectric effect and 15kV/cm field effects. Left direction shows application of electrical field directly after "No Piezoelectricity" case without including piezoelectric effect. Piezoelectric effect reorders P-states in the lower quantum dot and plays a critical role in determining the resonance at higher electrical field.

than ~30 nm will interact with each other through the piezoelectric potentials.

Past studies [26-29] have shown that the quadrupole nature of the piezoelectric effect significantly changes the splitting and orientation of electron P-states which are already oriented along diagonal directions due to atomistic interface and strain asymmetry. In figure 1(b), the optically active states in the experimental spectrum are identified as excited states ($E_3$ and $E_4$) in the range of applied bias. The next paragraph shows that considering piezoelectric effects is indeed critical in reproducing the experimental excitonic spectrum.

### *Quantitative Explanation of 'One' versus 'Two' Anti-Crossings*

Figure 5 presents a quantatively explanation of the presence of 'one' and 'two' anti-crossings. It shows a flow diagram in two directions: (Flow to right) First six electron states without piezoelectricity and without electrical field → with piezoelectricity but without electrical field → with piezoelectricity and with 15kV/cm electrical field, (Flow to left) First six electron states without piezoelectricity and without electrical field → without piezoelectricity and with 15kV/cm electrical field. It is critical to understand the arrangement of the electron states at 15kV/cm because at electrical field values from 15kV/cm to 23kV/cm, this will determine the resonance coupling. Since the first hole state remains in the upper quantum dot throughout the sweeping of the applied electrical field, it is not shown in figure 5. Piezoelectricity shifts the ground electron state $E_1$(U-S) only by a very small value of 0.59meV. The major contribution towards the shifting of the ground electron state comes from the externally applied bias. Hence the total shift in this state in either direction is approximately the same: Left=15.09meV, Right=15.01+0.59=15.6meV. The most significant effect of the piezoelectric effect is found to be on the $E_5(L-P_x)$ and $E_6(L-P_y)$ electron states. These states are not only shifted by a larger magnitude, but also their relative order is changed. Similar behavior has been observed in previous studies for a single quantum dot [25-27, 29], in which a strong piezoelectric effect has been shown to flip the order of the electron P-states. Here, the piezoelectric effect induces a shift of 0.7meV and 3.2meV in $E_5(L-P_x)$ and $E_6(L-P_y)$ respectively, followed by an additional electrical field induced shift of 7.7meV and 11.5meV respectively. This total shift of 3.2+11.5=14.7meV causes $E_6(L-P_y)$ to become $E_2(L-P_y)$ at 15kV/cm, which is below $E_3(U-S)$ on the same energy scale. A further increase of the electrical field beyond 15kV/cm will shift $E_3(U-S)$ to higher energies and $E_4(L-P_x)$ to lower energies. This will result in resonant coupling of these two states with one anti-crossing ($E_3 \rightarrow E_4$) in the emission spectra. In the left direction with no piezoelectricity included, there will be no additional shift of 3.2meV and 0.7meV in the lower quantum dot P-states. Mere electrical field induced shifts of 10.02meV and 10.11meV are not sufficient enough to push either of the two excited states below $E_2(U-S)$. This will result in two anti-crossings ($E_2 \rightarrow E_3 \rightarrow E_4$).

*Conclusions*:

In conclusion, a detailed systematic atomistic tight binding study of the experimental single quantum dot molecule geometry is presented, taking into account the long range strain, the atomistic interface asymmetry, and both the linear and the quadratic piezoelectric effects. The quantum dot molecule energy states are composed of individual single quantum dot states which can interact with each other depending on their energetic alignment. The external electrical field enables the tuning of relative alignment of molecular and atomic states which has been observed in experimental spectroscopy data [12]. Our calculations quantitatively reproduce the experimental spectra. The interband transition strengths are calculated to compare the relative intensities of the excitonic emissions and characterize them as optically active and inactive states. The calculated transition strengths closely follow the experimentally measured intensities and identify the excited electron (P-states) states as the optically active states in the range of the applied electrical field. The close quantitative agreement of our theoretical calculations with the experimental data [12] allows us to gain significant physical insight of a quantum dot molecule. We sweep the electrical field from zero to 21kV/cm and provide a spectroscopic mapping of the lower quantum dot electron states and the upper quantum dot hole excited states. Analysis of the observed sequence of the anti-crossings provides a technique to precisely probe the energy states of one quantum dot with the help of the energy states of the neighboring quantum dot. The direction of the applied electrical field determines the quantum dot whose electron/hole electronic states will be probed by increasing the electrical field magnitude. This technique can be used to 'reverse engineer' the geometry of the quantum dot molecule from the experimentally measured optical spectrum.

The strain and piezoelectric fields are both found to be long range effects. This imposes the need of a large size of simulation domain consisting of about 15 million atoms in the strain domain and about 9 million atoms in the electronic domain. Our calculations include both the first order and the second order piezoelectric fields. The quadrupole nature of the net piezoelectric effect along [110] and [$\bar{1}$10] directions significantly affects the splitting and orientation of the excited electron states. The NEMO 3-D based atomistic study shows that the net piezoelectric effect is critical in reproducing the experimentally observed optical transitions that are dominated by excited states such as devices with tuned with electrical fields. Continuum methods such as the effective mass model and k•p method do not take into account the atomistic nature of the InGaAs quantum dots and thus cannot correctly incorporate the strain and piezoelectric fields. The quantitative modeling of the experimental spectra [12] presented here allows us to extend our model for charged excitonic calculations. We are implementing the calculations of charged excitons in our model and a detailed analysis of a complex excitonic spectrum such as measured in the reference [37] will be presented in the future.

*Acknowledgements*:

Muhammad Usman acknowledges the Fulbright fellowship funding from the USA Department of States (Grant ID # 15054783). https://nanoHUB.org computational resources operated by the Network for Computational Nanotechnology (NCN), and funded by the National Science Foundation (NSF) are used for this work.

Matthias Tan acknowledges funding by the German Academic Exchange Service (DAAD) and a fellowship by the German National Merit Foundation (Studienstiftung des Deutschen Volkes). Gerhard Klimeck and Hoon Ryu acknowledge funding by the National Science Foundation (NSF). Shaikh Ahmed would like to acknowledge the ORNL/ORAU HPC Grant 2009.